\documentstyle[11pt,iau177,twoside,psfig2]{article}

\begin{document}


\title{Radio Emission Properties of Millisecond Pulsars}

\author{Michael Kramer}
\affil{Max-Planck-Institut f\"ur Radioastronomie, Bonn, Germany\\
University of Manchester, Jodrell Bank Observatory, UK}

\author{Kiriaki M. Xilouris}
\affil{Department of Astronomy, University of Virginia, Charlottesville, USA}

\begin{abstract}
We report the most recent progress in understanding the emission properties
of millisecond pulsars.
\end{abstract}

\vspace{-0.7cm}

\section{Introduction --- {\it Duo quum faciunt idem, not est idem.$^1$}}
\setcounter{footnote}{1}
\footnotetext{If two do the same, it is not the same.}

Through intensive research for almost two decades, it has been well
established, both in theory and observation, that millisecond pulsars
(MSPs) are the end product of mass accretion in binary systems. As
MSPs emerge in the radio universe having been given a second chance in
life, they are surrounded by magnetospheres which are several orders
of magnitude more compact than those of slower rotating
pulsars. Inferred magnetic fields close to the surface of MSPs are 3
to 4 orders of magnitude weaker than in normal pulsars while charges
at these regions experience an accelerating potential similar to that
of normal pulsars.  The impact of the different environment on the
emission process in MSP magnetospheres has been a question addressed
already shortly after the discovery of a first few such sources.

With the plethora of MSPs detected over the years, a significant
sample became available to us, enabling a better understanding of not
only MSPs (as radio sources and tools) but slower rotating (normal)
pulsars as well. In the following, we will concentrate on {\em recent}
progress, referring to Kramer et al.~(1998, Paper I) on spectra, pulse
shapes and beaming fraction; Xilouris et al.~(1998, Paper II) on
polarimetry of 24 MSPs; Sallmen (1998) and Stairs et al.~(1999) on
multi-frequency polarimetry; Toscano et al.~(1998) on spectra of
Southern MSPs; Kramer et al.~(1999b, Paper III) on multi-frequency
evolution; and Kramer et al.~(1999a, Paper IV) on profile
instabilities of MSPs; but see also the following contributions by
Kuzmin \& Losovsky and Soglasnov.

\vspace{-0.3cm}

\section{Single Pulses vs.~Average Profile Studies}

Single pulse observations still remain the only tool available to
address some fundamental questions listed below.  They are, however,
still technically challenging and the number of observations described
in the literature are scarce. In total, data for only three sources
describing 180 min of observations have been presented, i.e.~PSRs
B1937+21, B1534+12 and J0437--4715 (e.g.~Sallmen 1998, Cognard et
al.~1996, Jenet et al.~1998 and references therein). The results can
be summarized in the statement that based on the single pulses
studied, one cannot distinguish between a millisecond or slowly
rotating pulsar. More observations are required to further investigate
pulse fluctuations (e.g.~stabilization processes), the short-term
structure (e.g.~how it relates to microstructure) and in particular
the polarization characteristics in detail.  For the time being, we
investigate the wealth of information already provided by average
profile studies.

\begin{figure}[h]
\vspace{-0.2cm}
\centerline{\psfig{file=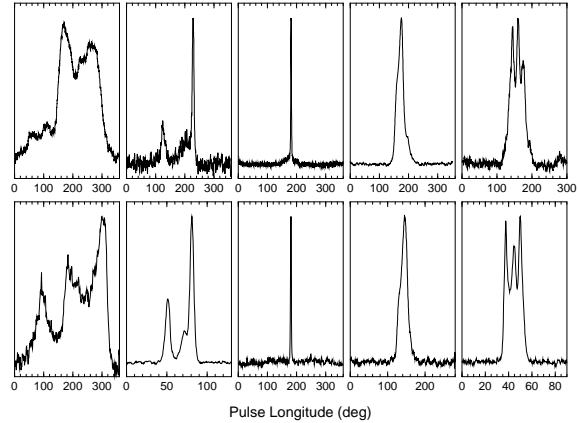,angle=-90,height=5.5cm} }
\caption{ Pulse profiles for a selected sample of MSPs and normal pulsars
(Paper I and EPN database). 
Note the similarity and make a guess which is which! See footnote$^2$
for the solution.}
\end{figure}

\vspace{-0.8cm}

\section{Flux Density Spectra and Radio Luminosity}

Prior to the investigations leading to Paper I it was commonly
believed that the spectra of millisecond pulsars were steeper than
those of normal pulsars.  We demonstrated in Paper I that the
distribution of spectral indices for MSPs is in fact not significantly
different, finding an average index of $-1.76\pm0.14$ (Paper III). The
initial impression was due to a selection effect, since the first MSPs
were discovered in previously unidentified steep spectrum sources, as
it was later pointed out by Toscano et al.~(1998).  Consequently, the
number of MSPs to be discovered in high-frequency surveys was
underestimated. The predictions for searches at frequencies as high as
5 GHz appear even more favourable in light of the latest results
presented in Paper III. These suggest that most spectra can be
represented by a simple power law, i.e.~clear indications for a
steepening at a few GHz as known from normal pulsars are not
seen. Extending the data to lower frequencies (see Paper III; Kuzmin
\& Losovsky, next contribution), evidence for spectral turn-overs were
not found.

\setcounter{footnote}{2}
\footnotetext{Upper row: MSPs (PSRs J0218+4218, J0621+1001, B1534+12,
J1640+2224, J1730$-$2304), lower row: normal pulsars (PSRs B1831$-$04,
B2045$-$16, B2110+27, B2016+28, B1826$-$17)}

Bailes et al.~(1997) pointed out that isolated MSPs are less luminous
than those in binary systems, pointing towards a possible relation
between radio luminosity and birth scenarios.  We have compared a
distance limited sample of normal pulsars and MSPs and came to a
similar result with the MSPs as a whole appearing as weaker sources than
normal pulsars.

\begin{figure}
\begin{tabular}{@{\hspace{-0.2cm}}c@{\hspace{-0.8cm}}c}
\psfig{file=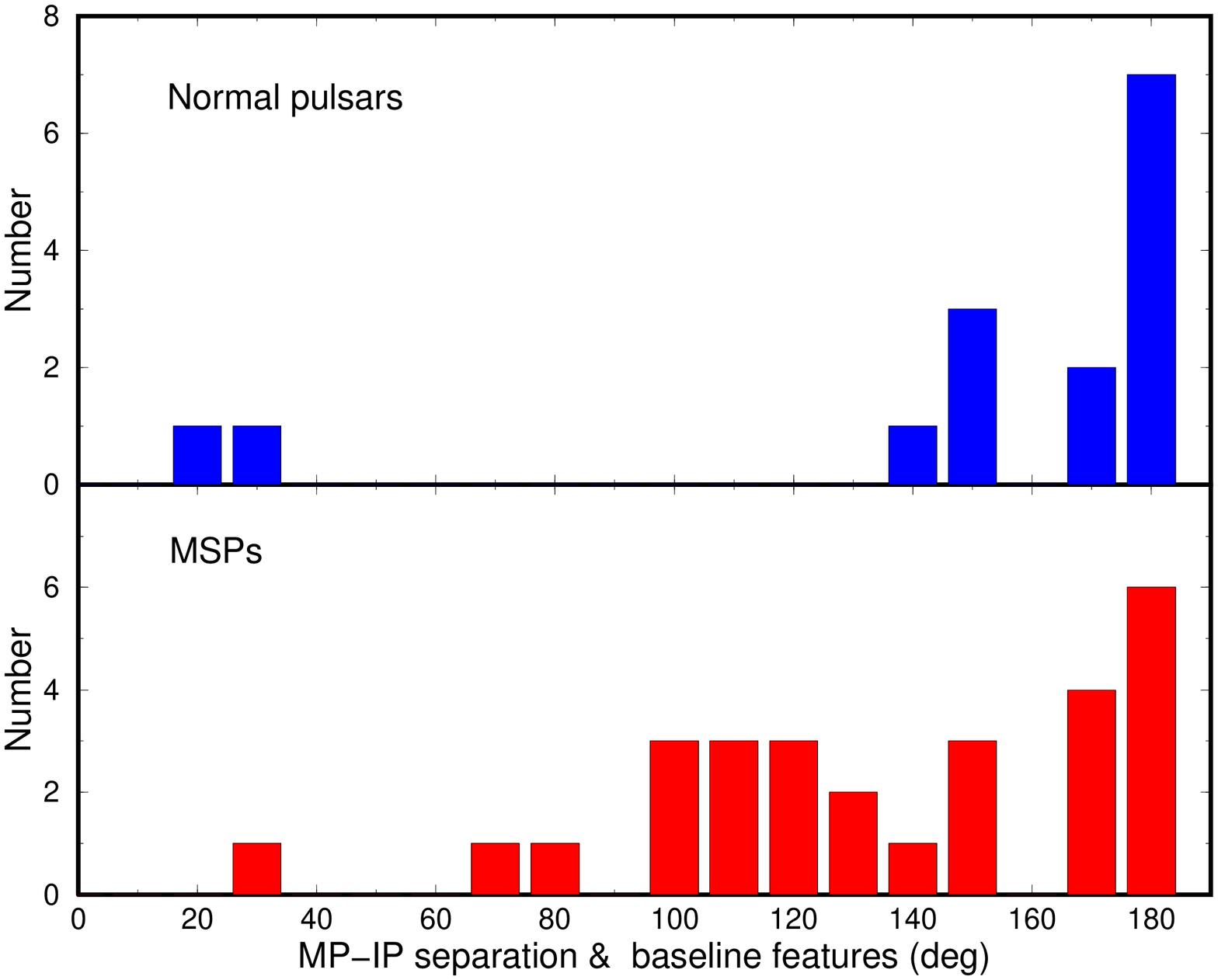,angle=-90,height=5.2cm} &
\psfig{file=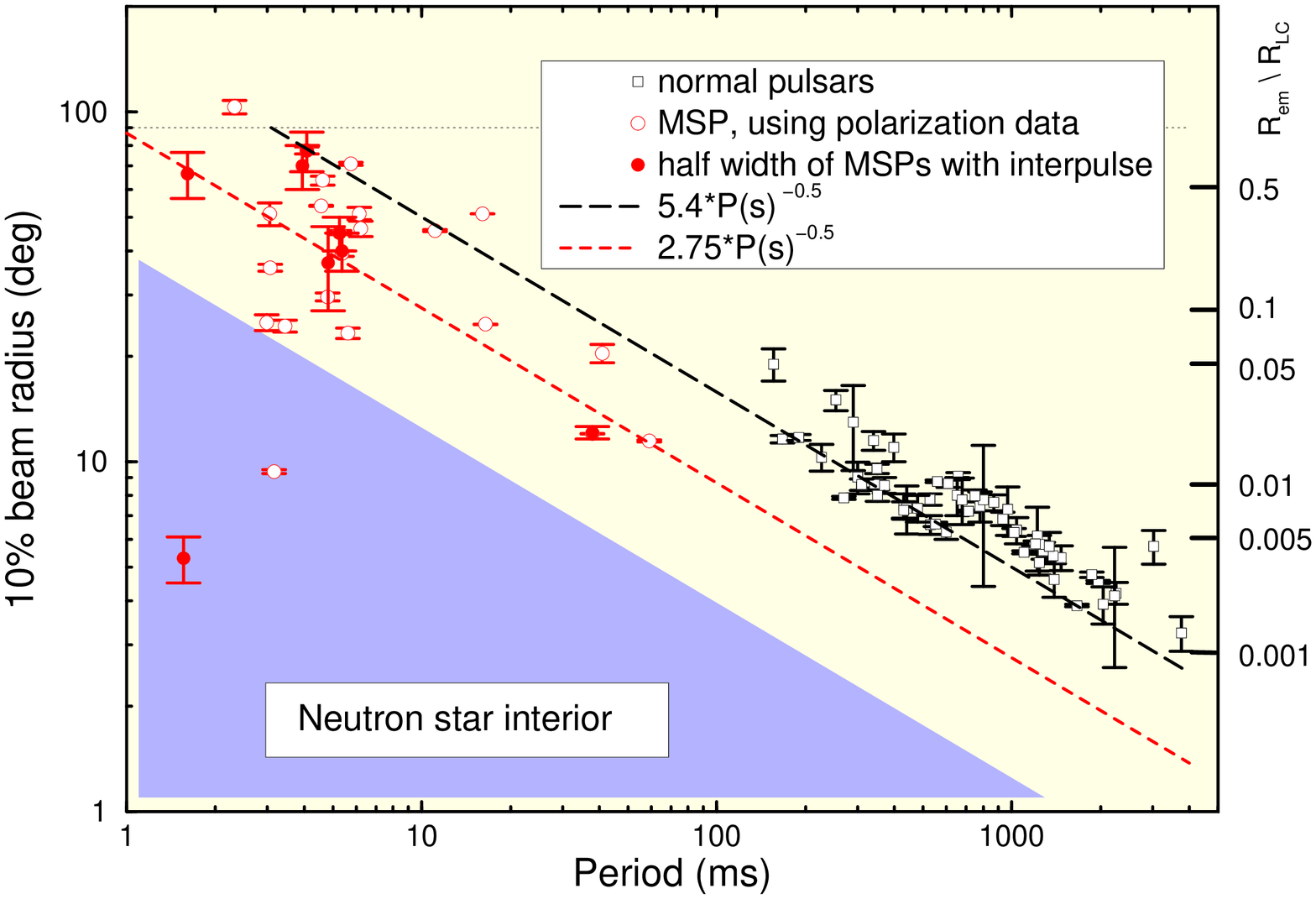,angle=-90,height=6cm} 
\end{tabular}
\caption{ a) Location of additional pulse features across the pulse period for
  normal pulsars and MSPs. b) The beam radius, $\rho$, for normal pulsars
  and MSPs. MSPs do not follow the scaling law of normal pulsars (here Gould
  1994) but their beaming fraction is much smaller. For MSPs with interpulses
an ``inner'' relationship is indicated.}
\end{figure}

\vspace{-0.3cm}

\section{Pulse Profiles -- Complexity, Interpulses and Beaming Fraction}

It was also believed that MSP profiles are more complex than those of
normal pulsars. Using a large uniform sample of profiles for fast and
slowly rotating pulsars, we showed in Paper I that the apparent larger
complexity is due to the (typically) larger duty cycle of MSPs.  As a
result we see ``blown-up'' profiles which make it easier to see
detailed structure. In fact, blown-up normal pulsar profiles show
very similar structure.  A quantitative proof is given in Paper I,
while Fig.~1 provides an illustration of this effect.

\begin{figure}
\begin{tabular}{cc}
\psfig{file=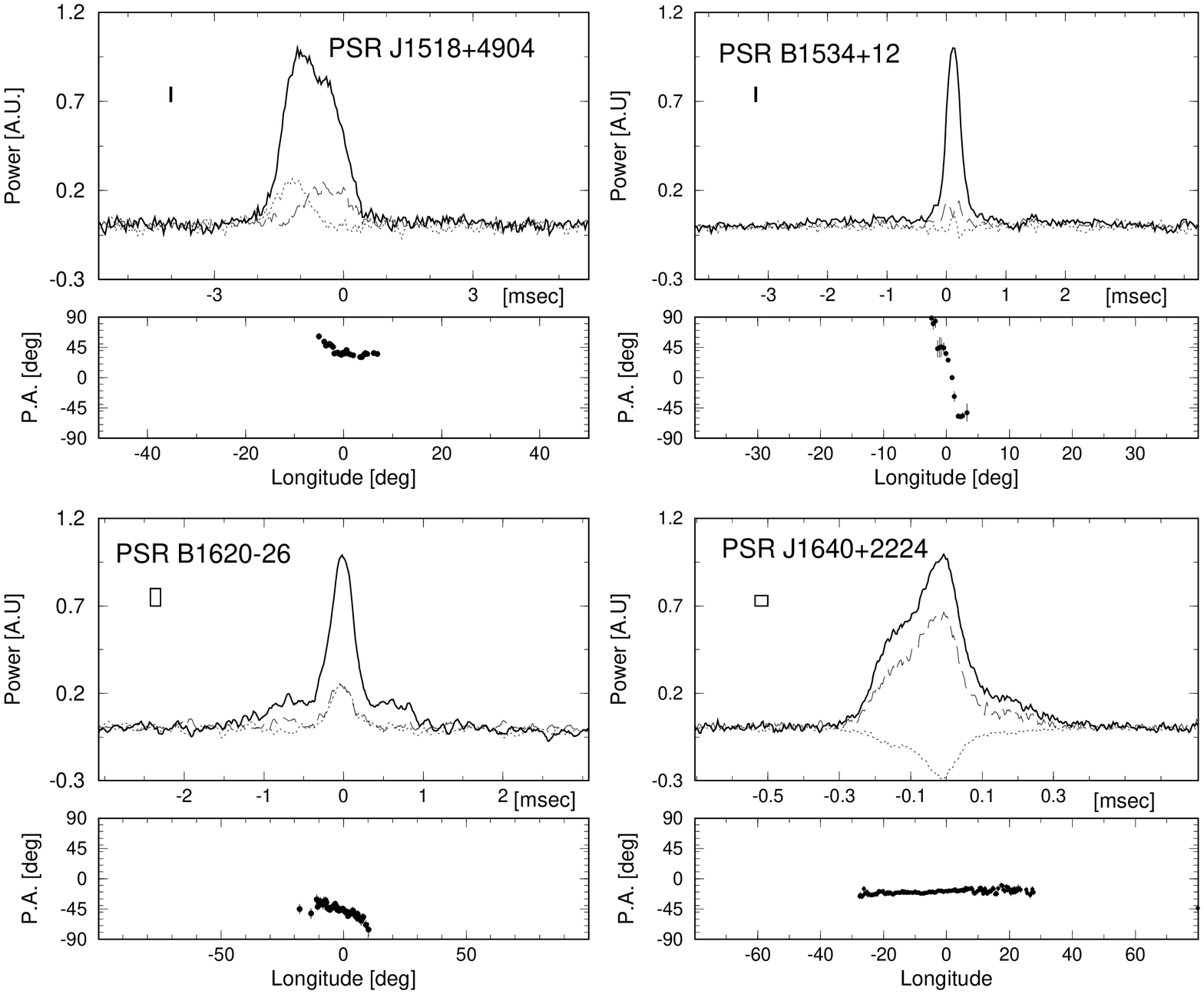,angle=-90,height=5cm,clip=} &
 \psfig{file=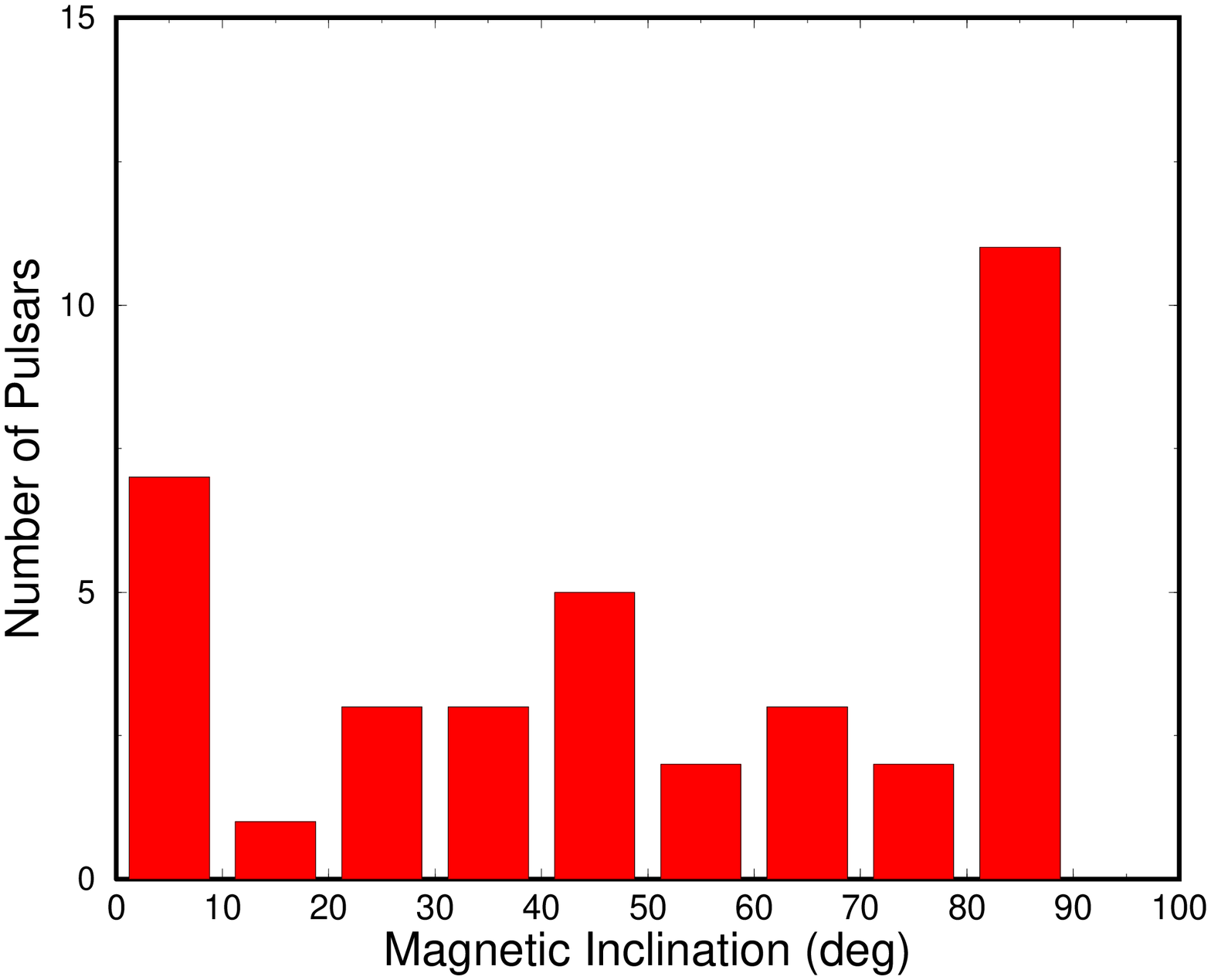,angle=-90,height=5.5cm} 
\end{tabular}
\caption{ a) PSR J1640+2224 as an example for a MSP exhibiting a flat PA swing,
b) distribution of magnetic inclination angles derived from RVM fits.}
\end{figure}

Despite this apparent similarity, there is a profound difference
betweent MSP profiles and those of normal pulsars! Additional pulse
features like interpulses, pre- or post-cursor are much more common
for MSPs. While only $\sim2$\% of all normal pulsars are known to show
such features, we detect them for more than 30\% of all (field)
MSPs. They also appear at apparently random positions across the pulse
period in contrast to normal pulsars (Fig.~2a). Their frequent
occurrence and location makes one wonder --- given the similarity of
the main pulse shapes otherwise --- whether these components are of
the same origin as the main pulse profile or whether other sources of
emission (e.g.~outer gaps) are responsible (see Paper II).  Other
possibilities involve an interpretation first put forward for some
young pulsars by Manchester (1996), who interpreted some interpulses
as the results of cuts through a very wide cone. This is an
interesting possibility also for MSPs, since their beam width appears
to be much smaller than predicted from the scaling law derived for
normal pulsars. The beam width of normal pulsars, $\rho$, i.e.~the
pulse width corrected for geometrical effects (see Gil et al.~1984),
follows a distinct $\rho \propto P^{-0.5}$-law (e.g.~Rankin 1993,
Kramer et al.~1994, Gould 1994). Using polarization information to
determine the viewing geometry and also applying statistical
arguments, we calculated $\rho$ (at a 10\% intensity level) for MSPs
in Paper I. We showed that they are not only much smaller than the
extrapolation of the known law to small periods, but that -- under the
assumption of dipolar magnetic fields -- the emission of some MSPs
seems to come even from within the neutron star --- a really
disturbing result! While we discuss the possibility of non-dipolar
fields and the used polarization information below, one explanation
would be that (perhaps below a critical period) the emission beam does
not fill the whole open field line region (``unfilled beam''). The
situation improves somewhat when we consider the additional pulse
features as regular parts of the pulse profile (Fig.~2b).  In fact,
those MSPs with interpulses may indicate an additional inner scaling
parallel to that known for normal pulsars, which could be a result of
unfilled beams.  We close this section by pointing out that the much
smaller beam width has consequences for population studies, which
usually utilize the $\rho-P$ scaling as found for normal pulsars. The
failure of this law leads to an overestimated beaming fraction and an
underestimation of the birth rate of recycled pulsars (see Paper I).

\begin{figure}
\begin{tabular}{c@{\hspace{-0.8cm}}c}
\psfig{file=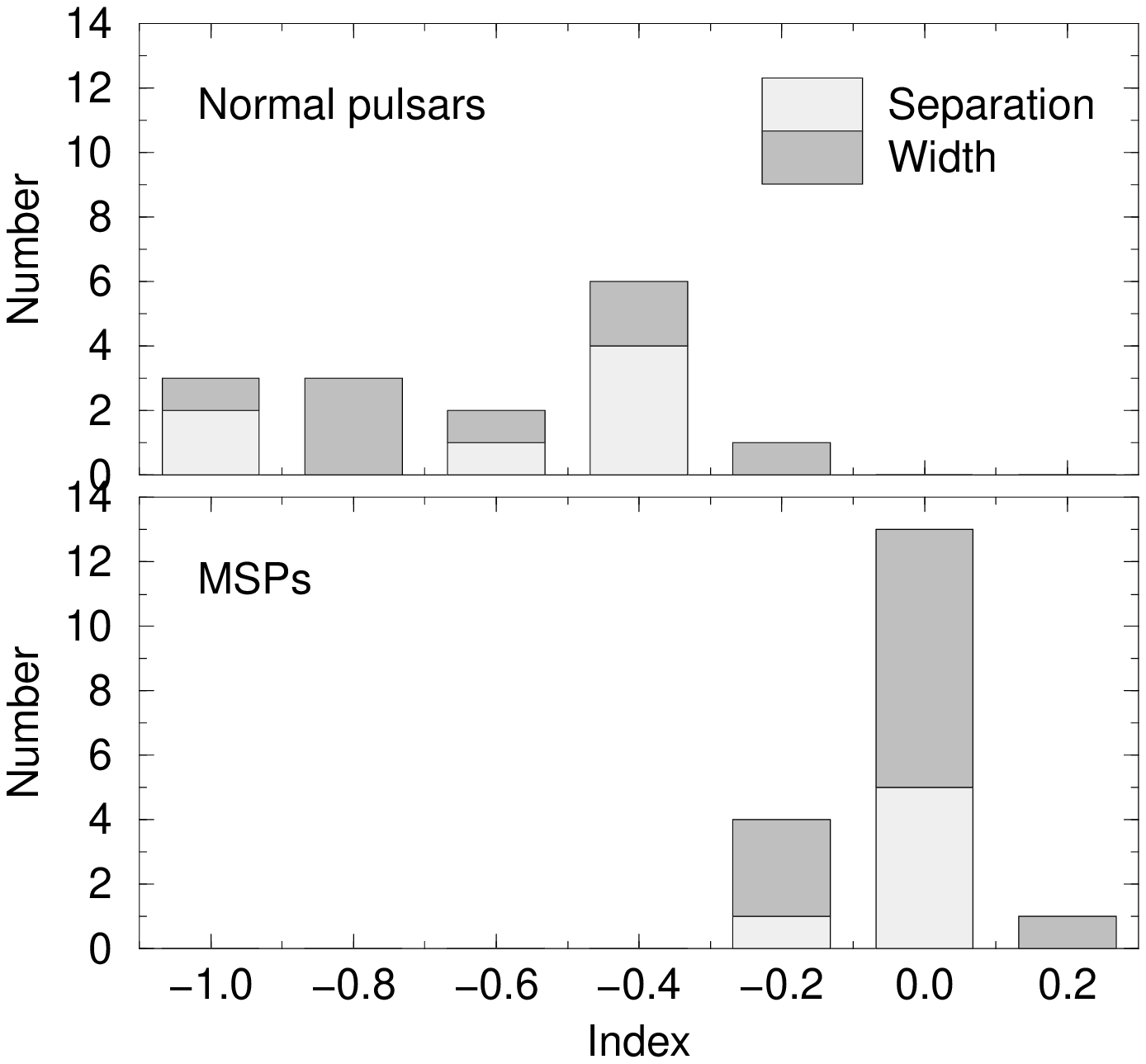,height=5cm} &
\psfig{file=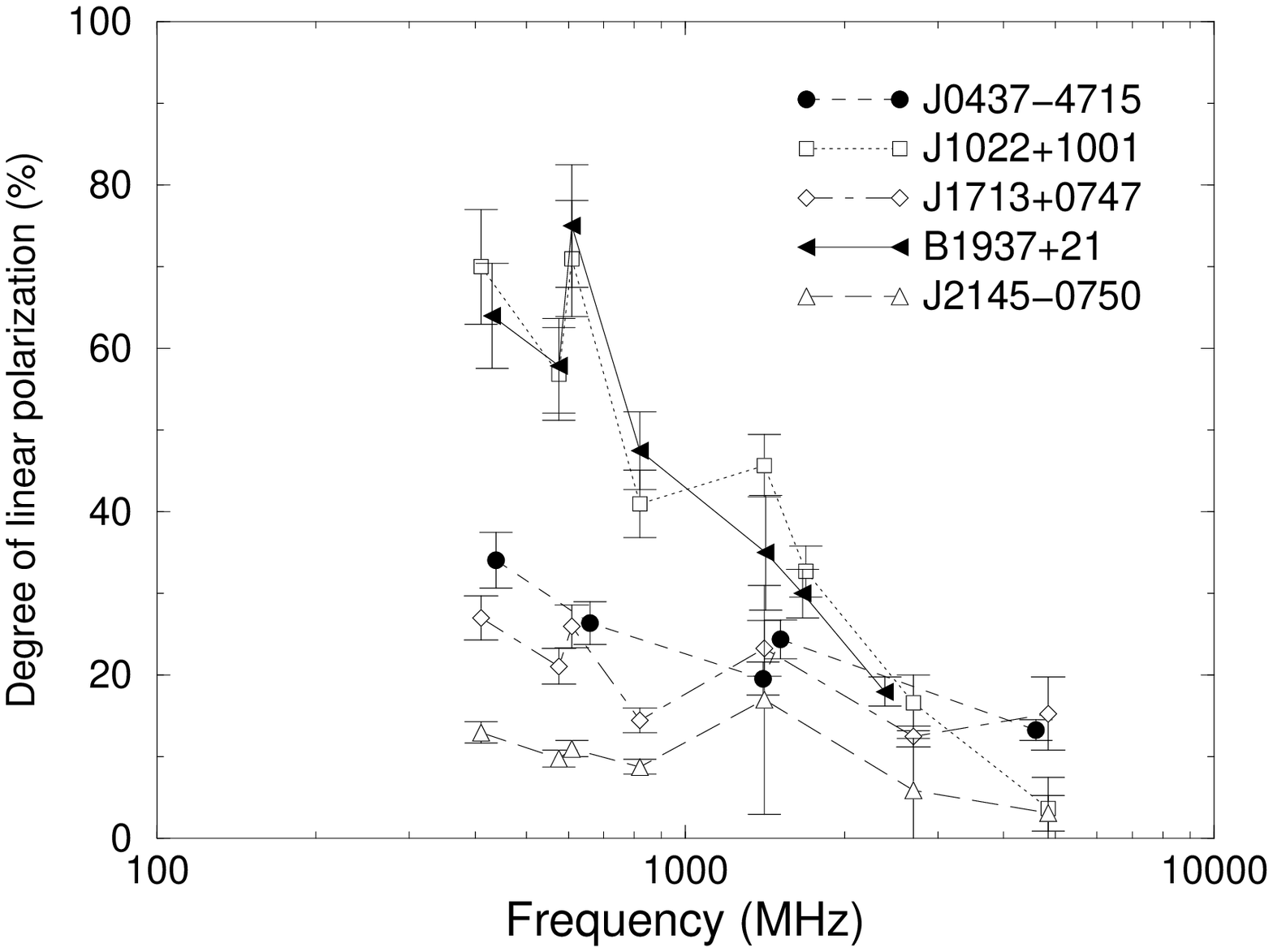,height=4.5cm} 
\end{tabular}
\caption{ a) Power law index of profile narrowing with frequency (see Paper III for details), b) degree of polarization for MSPs.}
\end{figure}

\vspace{-0.5cm}

\section{Polarization Properties}

The radio emission of MSPs shows all polarization features known from
normal pulsars, i.e.~circular polarization which is usually associated
with core components, linear polarization which is usually associated
with cone components, and also orthogonal polarization modes (see
Paper II, Sallmen 1998, Stairs et al.~1999). Despite the qualitative
similarities, the position angle (PA) swing is often strikingly
different.  While normal pulsars show typically a {\sf S}-like swing,
which is interpreted within the rotating vector model (RVM;
Radhakrishnan \& Cooke 1969), the PAs of many MSPs often appear flat
(see e.g.~Fig.~3a). This could be interpreted in terms of non-dipolar
fields, but Sallmen (1998) noted that larger beam radii lead to a
larger probability for outer cuts of the emission cones, i.e.~flatter
PA swings according to the RVM.  Although one should bear in mind the
limitations of the $\rho$-scaling law and another caveat
discussed later, this interpretation justifies the geometrical
interpretation of the data, which is supported by the results of
Hibschman (these proceedings).  Magnetic inclination angles derived
from RVM fits are important for binary evolution models and
determinations of the companion mass (Fig.~3b).

\vspace{-0.3cm}

\section{Frequency Evolution}

The radio properties of normal pulsars show a distinct frequency
evolution, i.e.~with increasing frequency the profile narrows, outer
components tend to dominate over inner ones, and the emission
depolarizes. The emission of MSPs, which at intermediate frequencies
tends to be more polarized than that of normal pulsars (Paper II),
also depolarizes at high frequencies (Fig.~4b; Paper III).
Simultaneously, the profile width hardly changes or remains constant
(see Fig.~4a, Paper III; Kuzmin \& Losovsky, these proceedings).  This
puts under test attempts to link both effects to the same physical
origin (i.e.~birefringence). In fact, many profiles also exhibit the
same shape at all frequencies, while others evolve in an unusual way,
i.e.~the spectral index of inner components is not necessarily
steeper, so that a systematic behaviour as seen for normal pulsars is
hardly observed. This can be understood in terms of a compact emission
region, an assumption further supported by a simultaneous arrival of
the profiles at all frequencies. We emphasize that we have not
detected any evidence for the existence of non-dipolar fields in
the emission region (Paper III).

\vspace{-0.3cm}

\section{Profile and Polarization Instabilities}

The amazing stability with time of MSP profiles has enabled high
precision timing over the years. However, in Paper IV we discussed the
surprising discovery that a few MSPs do show profile changes caused by
an unknown origin.  The time scales of these profile instabilities are
inconsistent with the known mode-changing. In particular, PSR
J1022+1001 exhibits a narrow-band profile variation never seen before
(Paper IV), which could, however, be the result of magnetospheric
scintillation effects described by Lyutikov (these proceedings). With
the pulse shape the polarization usually changes as well, and hence
this effect is possibly related to phenomena which we discovered in
Paper II. Some pulsars like PSR J2145--0750 (Paper II) or PSR
J1713+0747 (Sallmen 1998) show occasionally a profile which is much
more polarized than usual. In the case of PSR J2145--0750, the PA
also changes from some distinct (though not {\sf S}-like) swing to some
very flat curve. This is a strong indication that some of the flat PA
swings discussed above may not be of simple geometrical origin alone.

\section{Summary -- MSPs in 2000 and Beyond}

While we have had to be necessarily brief in reviewing MSP properties,
we direct the interested reader to the extensive studies of MSPs
presented in the quoted literature. We summarize here our point of
view: MSPs emit their radio emission by the same mechanism as normal
pulsars.  Some distinct differences may originate from the way they
were formed, but most observed features can be explained by very
compact magnetospheres. Our data can be explained without any need to
invoke deviations from dipolar field lines, although a large number of
open questions remain. We need more polarization information at higher
frequencies and, in particular, single pulse studies.  These will
allow us to study the formation of the profile and its stability, to
see if the additional pulse features are distinct from the main pulse,
and how the polarization modes behave under the magnifying glass of
the blown-up MSP profiles. There are exciting years to come!

\acknowledgements We are very grateful to all the people involved in the
studies of MSPs at Bonn, i.e.~Don Backer, Fernando Camilo, Oleg Doroshenko, 
Alexis von Hoensbroech, Axel Jessner, Christoph Lange, Dunc Lorimer, 
Shauna Sallmen, Norbert Wex, Richard Wielebinski and Alex Wolszczan.

\end{document}